\documentclass[aps,prd,amsmath,amssymb]{revtex4}

\usepackage{graphicx}
\usepackage{dcolumn}
\usepackage{bm}
\usepackage{amsmath}
\usepackage{epstopdf}
\usepackage{amsfonts}
\usepackage{amssymb}
\usepackage{epsfig}
\usepackage{tabularx}
\usepackage{color}
\usepackage{soul}

\usepackage[colorlinks=true,citecolor=blue,urlcolor=magenta,breaklinks]{hyperref}

\newcommand{\be}{\begin{equation}}
\newcommand{\en}{\end{equation}}
\newcommand{\bea}{\begin{eqnarray}}
\newcommand{\ena}{\end{eqnarray}}

\begin{document}

\title{Slowly rotating dark energy stars}

\author{
Grigoris Panotopoulos,{${}^{a}$,${}^{b}$
\footnote{
\href{mailto:grigorios.panotopoulos@ufrontera.cl}{grigorios.panotopoulos@ufrontera.cl} 
}
{\'A}ngel Rinc{\'o}n,{${}^{c}$
\footnote{
\href{mailto:aerinconr@academicos.uta.cl}{aerinconr@academicos.uta.cl} 
}
}
Il{\'i}dio Lopes{${}^{a}$
\footnote{
\href{mailto:ilidio.lopes@tecnico.ulisboa.pt}{ilidio.lopes@tecnico.ulisboa.pt} 
}
}
}
}

\address{
${}^a$ Centro de Astrof{\'i}sica e Gravita{\c c}{\~a}o-CENTRA, Instituto Superior T{\'e}cnico-IST, Universidade 
de Lisboa-UL, Av. Rovisco Pais, 1049-001 Lisboa, Portugal. \\
${}^b$ Departamento de Ciencias F{\'i}sicas, Universidad de La Frontera, Casilla 54-D, 4811186 Temuco, Chile. \\
${}^c$ Sede Esmeralda, Universidad de Tarapac\'a, Avenida Luis Emilio Recabarren 2477, Iquique, Chile. 
}

\begin{abstract}
We study isotropic and slowly-rotating stars made of dark energy adopting the extended Chaplygin equation-of-state.
We compute the moment of inertia as a function of the mass of the stars, both for rotating and non-rotating objects. 
The solution for the non-diagonal metric component as a function of the radial coordinate for three different star masses
is shown as well. We find that i) the moment of inertia increases with the mass of the star, ii) in the case of non-rotating objects the moment of inertia grows faster, and iii) the curve corresponding to rotation lies below the one corresponding to non-rotating stars. 
\end{abstract}

\maketitle

%%%%%%%%%%%%%%%%%%%%%%%%%%%%%%%%%%%%%
\section{Introduction}\label{Intro}
%%%%%%%%%%%%%%%%%%%%%%%%%%%%%%%%%%%%%

The origin of dark energy (DE), i.e. the fluid component that dominates the current cosmic acceleration, \cite{SN1,SN2,turner}, is still a mystery, while at the same understanding its nature comprises one of the biggest challenges in modern theoretical Cosmology. It is well-known that according to the cosmological equations within Einstein's General Relativity \cite{GR}, a Universe consisting of radiation and non-relativistic matter only cannot expand at an accelerating rate. On the contrary, a non-vanishing (and positive) cosmological constant \cite{einstein,carroll} has been proven to be the most economical model in an overall excellent agreement with a wealth of available observational data.

\smallskip

The $\Lambda$CDM model, based on collisionless dark matter and a positive cosmological constant, despite its success, does not come without problems. In modern times the community is facing a couple of puzzles related on the one hand to the cosmological constant problem \cite{zeldovich,weinberg} and on the other hand to the Hubble tension. To be more precise, regarding the value of the Hubble constant $H_0$, there is nowadays a disagreement between high red-shift CMB data and local measurements at low red-shift data, see e.g. \cite{tension,tension1,tension2,tension3}. The value of the Hubble constant determined by the PLANCK Collaboration \cite{planck1,planck2}, $H_0 = (67-68)~\text{km/(Mpc  sec)}$, is found to be lower than the value obtained by local measurements, $H_0 = (73-74)~\text{km/(Mpc sec)}$ \cite{hubble,recent}. This disagreement  might call for new physics \cite{newphysics}.

\smallskip

For those reasons, as was to be expected, a plethora of several different dark energy models have been proposed and studied 
over the years as possible alternatives to the $\Lambda$CDM model. 
Quite generically, all dark energy models are classified into two broad classes. On the one hand, there is a family of models related to alternative/modified theories of gravity, where new correction terms appear to GR at cosmological scales. And on the other hand, in another family of models, we introduce a new dynamical field with an equation-of-state (EoS) parameter $w < -1/3$.
 In the first class of models, called geometrical DE, one finds for instance $f(R)$ theories of gravity \cite{mod1,mod2,HS,starobinsky}, brane-world models \cite{langlois,maartens,dgp} and Scalar-Tensor theories of gravity \cite{BD1,BD2,leandros,PR}, while in the second class, called dynamical DE, one finds models such as quintessence \cite{DE1}, phantom \cite{DE2}, quintom \cite{DE3}, tachyonic \cite{DE4} or k-essence \cite{DE5}. For an excellent review on the dynamics of dark energy see e.g. \cite{copeland}. Of particular interest is the Chaphygin gas dark energy model and its generalization \cite{Chaplygin1,Chaplygin2}, which unifies non-relativistic matter with the cosmological constant introducing a single fluid with an equation-of-state $p=-B^2/\rho^\omega$, where $B,\omega$
are positive constant parameters, and $\omega$ takes values in the range $0 < \omega \leq 1$.

\smallskip

Abandoning the cold dark matter paradigm, in which dark matter is collisionless, models where dark matter exhibits self-interactions have been proposed as an attractive and elegant solution to the dark matter crisis at short (galactic) scales \cite{Tulin}. In such a scenario it is not contrived to imagine objects made entirely of self-interacting dark matter, see e.g. \cite{DMS1,DMS2,DMS3,DMS4}. Similarly, given that the current cosmic acceleration calls for dark energy, in a few recent works the authors considered the possibility that spherical configurations made of dark energy, or more generically exotic matter, just might exist \cite{exotic,paperbase1,paperbase2}.  The possibility of obtaining interior solutions of relativistic stars and gravitationally bounded configurations in different astrophysical contexts (such as anisotropic matter, carrying a net electric charge, non-conventional theories of gravity etc) is a very exciting proposal; there is a vast amount of publications in the literature.
For a partial list see e.g.~\cite{herre,harko,a2,a4,a5,a6,a7,a8,a9,a10,a11,a12,a13,a15,a16,a17,a19,a20,a21,a22,Panotopoulos:2021obe,Panotopoulos:2020kgl,Bhar:2020ukr,Abellan:2020dze,Tello-Ortiz:2020nuc,Tello-Ortiz:2020svg,Panotopoulos:2020zqa,Panotopoulos:2019zxv,Gabbanelli:2018bhs,Panotopoulos:2021cxu,Panotopoulos:2021sbf,Moraes:2021lhh,Tello-Ortiz:2020euy,Tello-Ortiz:2019gcl} and references therein.

\smallskip

In the present work we propose to study non-rotating dark energy stars, extending the works of \cite{paperbase1,paperbase2}, where non-rotating stars were considered, with isotropic matter assuming an extended Chaplygin EoS \cite{ExtCh1,ExtCh2,Extra1,Extra2,Extra3} of the form $p=-B^2/\rho + A^2 \rho$, where a barotropic term is added to the standard Chaphygin equation-of-state with $\omega=1$.

\smallskip

Our plan in the present article is the following: After this introduction, in the next section we briefly review the structure equations for non-rotating relativistic stars. In section 3 we add a non-vanishing angular momentum, we obtain the solutions, and we show and discuss discussing our main numerical results. Finally, we close our work with some concluding remarks in the last section. We adopt the mostly positive metric signature, $(-,+,+,+)$, and we work in geometrical units where the speed of light in vacuum as well as Newton's constant are set to unity, $c=1=G$.

%%%%%%%%%%%%%%%%%%%%%%%%%%%%%%%%%%%%%%%%%%%%%%%%%%%%%%%%%%%%%%%%%%%%
\section{Hydrostatic equilibrium of non-rotating relativistic stars}
%%%%%%%%%%%%%%%%%%%%%%%%%%%%%%%%%%%%%%%%%%%%%%%%%%%%%%%%%%%%%%%%%%%%

Here we shall briefly review the set of structure equations \cite{Tolman,OV} required to describe interior solutions 
of non-rotating relativistic stars within GR.

For a stationary, axially symmetric metric in Schwarzschild-like coordinates, $(t,r,\theta,\phi)$, and using the metric tensor, we can adopt the following ansatz in the slowly-rotating approximation:
\begin{equation}
\mathrm{d}s^2 = -e^{2 \nu(r)} \mathrm{d}t^2 + A(r) \mathrm{d}r^2 + r^2 (d \mathrm{\theta^2} + \mathrm{sin^2 \theta \: d \phi^2}),
\end{equation}
where $e^{2\nu(r)}$ and $A(r) \equiv e^{2\lambda(r)} $ are the metric potentials depending on the radial coordinate only.
To simplify the treatment we shall split the process in two steps: first, we will obtain the usual TOV equations for the non-rotating case in this section, and then we shall consider first order corrections due to a slow rotation, see next section.

\textbf{Step $\#$ 1:} For the non-rotating case, as usual we introduce for convenience the mass function, $m(r)$, 
defined by
\begin{equation}
A(r)^{-1} \equiv 1 - \frac{2 m(r)}{r}
\end{equation}
Moreover, if matter content is modeled as a perfect fluid it will be characterized by a stress-energy tensor of the form 
\begin{equation}
T_\nu ^\mu = \text{diag}(-\rho, p, p, p)
\end{equation}
with $\rho$ being the energy density and $p$ being the pressure. In Einstein's field equations, the $tt$ and $rr$ field equations yield
\begin{eqnarray}
m'(r) & = & 4 \pi r^2 \rho(r) \\
\nu'(r) & = & \frac{m(r)+4 \pi r^3 p(r)}{r^2 (1-2m(r)/r)}
\end{eqnarray}
respectively, where a prime denotes differentiation with respect to the radial coordinate $r$. Finally, instead of the angular field equations, equivalently one is allowed to make use of the conservation of energy, which reads
\begin{equation}
p'(r) = - [\rho(r) + p(r)] \nu'(r)
\end{equation}
Therefore, one obtains the usual TOV equations \cite{Tolman,OV} 
\begin{eqnarray}
m'(r) & = & 4 \pi r^2 \rho(r) \\
p'(r) & = & - [\rho(r) + p(r)] \: \frac{m(r)+4 \pi r^3 p(r)}{r^2 (1-2m(r)/r)} \\
\nu'(r) & = & - \frac{p'(r)}{\rho(r)+p(r)}
\end{eqnarray}
%
%%%%%%%%%%%%%%%%

The above equations need to be supplemented by an EoS such as $p(\rho)$ or $\rho(p)$, which will be discussed in the next section. 
Besides, we impose the following conditions both at the center of the star,  $r=0$
\begin{align}
m(0)=0,  \hspace{1cm}  \text{and}  \hspace{1cm}
p(0)=p_c,
\end{align}
and at the surface of the star $r=R$
\begin{align}
m(R)=M, \hspace{1cm}  \text{and}  \hspace{1cm}
p(R)=0. 
\end{align}
The latter is used to compute the radius, $R$, and the mass, $M$, of the object.
Moreover,  the corresponding metric potential $\nu$ is also determined by integrating Eq (9) plus the condition at 
the surface of the star, i.e., 
\begin{equation}
e^{2 \nu(R)} = 1 - \frac{2M}{R}.
\end{equation}
Therefore, the solution for $\nu(r)$ is given by
\begin{equation}
\nu(r) = \nu(R) - \int_R^r \frac{p'(z)}{p(z)+\rho(z)}
\end{equation}
where its surface value is given by
\begin{align}
\nu(R)=\frac{1}{2} \ln\left(1-\frac{2M}{R}\right).
\end{align}
%

%%%%%%%%%%%%%%%%%%%%%%%%%%%%%%%%%%%%%%%%%%%%%%%%%%%%%%%
\section{Stellar modeling of rotating relativistic stars}
%%%%%%%%%%%%%%%%%%%%%%%%%%%%%%%%%%%%%%%%%%%%%%%%%%%%%%%

\textbf{Step $\#$ 2:} To study non-rotating stars \cite{Gourgoulhon:2010ju,Paschalidis:2016vmz}, for the interior 
problem we make for the metric tensor the following ansatz
\begin{align}
\begin{split}
\mathrm{d}s^2 = &-e^{2 \nu(r)} \mathrm{d}t^2 + A(r) \mathrm{d}r^2 + r^2 (d \mathrm{\theta^2} + \mathrm{sin^2 \theta \: d \phi^2})
\\
&-2 \omega(r,\theta) r^2 \sin^2 \theta \mathrm{d}\phi \mathrm{d}t,
\end{split}
\end{align}
where now there is a non-diagonal metric component to account for the rotation of the object.

We shall now consider the differential equation for the $t_\phi$ component:
\begin{align}
R_{\phi}^t = 8 \pi T_{\phi}^t
\end{align}
and we will obtain the first order (linear) contributions only. 
At this point we recall that $\Omega$ is the angular velocity of the fluid (which is a constant for an uniformly rotating configuration) as seen by an observer at rest at some point 
$(t,r,\theta,\phi)$ in the fluid, whereas $\omega(r,\theta)$ is the angular velocity acquired by an observer
falling freely from infinity calculated to first order in $\Omega$. Thus, $\Omega - \omega$  give us the coordinate angular velocity of the fluid element at $(r,\theta)$ seen by the freely falling observer.

To obtain the contribution $T^t_{\phi}$, we first consider the normalization condition $u^{\mu}u_{\mu}=-1$, with
\begin{align}
u^t &= \sqrt{-(g_{tt} - 2\Omega g_{t \phi} + \Omega^2 g_{\phi \phi})}
\\
u^r &= 0
\\
u^{\theta} &= 0
\\
u^{\phi} &= \Omega \sqrt{-(g_{tt} - 2\Omega g_{t \phi} + \Omega^2 g_{\phi \phi})}
\end{align}
and the first order contributions is then 
\begin{align}
T^t_{\phi} &= (\rho + p)u^t u_{\phi} 
=  (\rho + p) e^{-2\nu}(\Omega - \omega)r^2\sin^2 \theta. 
\end{align}
We define the following convenient  quantity
\begin{align}
\tilde{\omega}(r,\theta) &= \Omega - \omega(r,\theta),
\end{align}
and writing down the first-order contribution of the Einstein field equation, for which we obtain
\begin{align} \label{eqperturb}
\begin{split}
& \frac{1}{r^4} \frac{\partial}{\partial r}
\Bigg[
  A^{-1/2} e^{-\nu} r^4 \frac{\partial \tilde{\omega}}{\partial r}
\Bigg] 
+
\frac{A^{1/2} e^{-\nu}}{r^2 \sin^3 \theta} \times
\\
&
\frac{\partial}{\partial \theta} 
\Bigg[
\sin^3 \theta \frac{\partial \tilde{\omega}}{\partial \theta} 
\Bigg]
=
 16 \pi (\rho + p)A^{1/2}e^{-\nu} \tilde{\omega} = 0.
\end{split}
\end{align}
As pointed out by Hartle \cite{Hartle:1967he}, we can use the method of separation of variables with the help of an expansion in vector spherical harmonics. Thus, taking into account the following expansion
\begin{align}
\tilde{\omega} (r, \theta) &=  \sum_{l=1}^{\infty} \tilde{\omega}_l (r)   \Bigg( - \frac{1}{\sin \theta} \frac{dP_l}{d\theta}  \Bigg) 
\end{align}
the radial functions $\tilde{\omega}_l$ must satisfy
\begin{align}
\begin{split}
& \frac{1}{r^4} \frac{d}{dr}
\Bigg[
 A^{-1/2} e^{-\nu} r^4  \frac{d \tilde{\omega}_l}{d r}
\Bigg] 
-
\frac{A^{1/2} e^{-\nu}}{r^2}(l(l+1) - 2)
 \tilde{\omega}_l
\\
&= 
16 \pi (\rho + p)  A^{1/2} e^{-\nu} \tilde{\omega}_l.
\end{split}
\end{align}
For the exterior solution in asymptotically flat space-times we obtain 
\begin{align}
\tilde{\omega}_l \rightarrow \alpha r^{-(l+2)} + \beta r^{l-1}
\end{align}
and  for $r \rightarrow \infty$,  we have
\begin{align}
\tilde{\omega}_l \rightarrow  - 2J r^{-3}+ \Omega.
\end{align}
We  conclude that $\tilde{\omega}_l = 0$ for $l \geq 2$, whereas for $l=1$ we obtain a simpler equation \cite{Staykov:2014mwa,PLPulsars}
\begin{align}
\begin{split}
 \frac{1}{r^4} \frac{d}{dr}
\Bigg[
 A^{-1/2} e^{-\nu} r^4  \frac{d \tilde{\omega}}{d r}
\Bigg] 
&= 
16 \pi (\rho + p)  A^{1/2} e^{-\nu} \tilde{\omega},
\end{split}
\end{align}
while the boundary conditions are given by \cite{Staykov:2014mwa,PLPulsars}
\begin{align} \label{eqomega}
\frac{d\tilde{\omega}}{dr}(0)=0,  \hspace{1cm}  \text{and}  \hspace{1cm}
\lim_{r \rightarrow \infty} \tilde{\omega} = \Omega.
\end{align}
We can define the moment of inertia, $I$, of the star as follows
\begin{align}
I \equiv \frac{J}{\Omega},
\end{align}
with $J$ being the angular momentum of the star.
Finally, utilizing \eqref{eqomega} plus the asymptotic form of $\tilde{\omega}$ we obtain the following expression for
the moment of inertia of a rotating star \cite{Staykov:2014mwa,PLPulsars}
\begin{align}
I &= \frac{8 \pi}{3} \int_0^{R} (\rho + p)e^{-\nu}A^{1/2} r^4 \left( \frac{\tilde{\omega}}{\Omega} \right) dr.
\end{align}

\subsection{Equation-of-state}

To close the system of differential equations, we must include an EoS for the matter content. 

In the following, we shall  adopt the extended Chaplyin EoS, which has been considerably used in a cosmological context and, not long ago, in stellar modeling of compact stars. The basic form of the above relation acquire the simple form $p=-\hat{B}/\rho$ where $\rho$ is the energy-density, $p$ is the pressure and $\hat{B}$ is a positive constant with units of $\text{length}^{-4}$. 
Albeit the previous relation allows us to get some insight about the physics, such expression is inconsistent with observational data, that is the reason why such equation was generalized \cite{r19}, namely $p=-\hat{B}/\rho^{\omega}$ for which $\omega $ is such that $0 < \omega \leq 1 $. Subsequently, another generalization was obtained in \cite{r20,r21} taking into account viscosity. Finally,  in Ref \cite{r22} such equation is improved by the inclusion of an additional term, to obtain
\begin{equation}\label{eq3}
p=\hat{A}\rho-\frac{\hat{B}}{\rho^{\omega}},
\end{equation}
where $\hat{A}$ a positive numerical value. The generalized Chaplygin EoS has been considerably used in different context, for example: 
i) compact stars in the framework of $f(T)$ gravity theory \cite{r26,r27},
ii)  wormhole geometries \cite{r28},
iii) charged anisotropic fluid objects \cite{r25},
iv)  and $5$-dimensional cosmology \cite{r29}.
Let us reinforce that the Chaplygin relations have been significantly used in the cosmological scenario. The latter can be understood because Chaplygin-like EoS correctly describes dark matter and dark energy. 
In the present paper, however, we will consider the extended Chaplygin EoS as follows:
\begin{align}
p=A^2 \rho - \frac{B^2}{\rho}.
\end{align}

\subsection{Moment of inertia: numerical results}

Here we obtain the numerical solution, and we present and discuss our main results. We shall consider three concrete models 
as follows:
\begin{equation}
A = \sqrt{0.4}, \ \quad B=0.23 \times 10^{-3}/km^2 \; \; \; (\textrm{model I})
\end{equation}
\begin{equation}
A = \sqrt{0.425}, \ \quad B=0.215 \times 10^{-3}/km^2 \; \; \; (\textrm{model II})
\end{equation}
\begin{equation}
A = \sqrt{0.45}, \ \quad B=0.2 \times 10^{-3}/km^2 \; \; \; (\textrm{model III}).
\end{equation}
Note that when the pressure vanishes at the surface of the star, the energy density 
takes the surface value $\rho_s = B/A$.

\smallskip

The numerical values assumed here lead to masses and radii of stars similar to those of neutron stars and strange quark stars, namely a mass $M \sim M_{\odot}$ and $R \sim 10~km$. The very same numerical values of $A,B$ were considered in \cite{paperbase2}, and they are comparable to the ones considered in \cite{paperbase1} where anisotropic fluid spheres were 
studied.

\smallskip

Once the EoS is known, we integrate the structure equations numerically imposing the initial conditions at the
origin as well as the matching conditions at the surface of the star. We thus compute all the unknown quantities (mass
function, pressure etc) as a function of the radial coordinate, and also the properties of the star, such as mass, radius, factor of compactness etc. Those were studied and discussed in our previous work. Here we are interested in the moment of inertia and on the effect of a non-vanishing rotation speed.

To demonstrate how observational data can validate such a class of models, we compute here the angular momentum of a pulsar with a known mass and frequency for the three EoS models discussed above. The pulsar J1738+0333 in one that we studied in a previous work (see \cite{PLPulsars} and references therein): this is rotating compact object with a mass at 1.47 solar masses, and a pulsar frequency at 170.9 Hz. For the angular momentum of this object within the three EoSs considered here, we obtain 
\begin{equation}
J_A = 1.77 \times 10^{41} kg \: m^2 / sec \; \; \; (\textrm{model I})
\end{equation}
\begin{equation}
J_B = 1.89 \times 10^{41} kg \: m^2 / sec  \; \; \; (\textrm{model II})
\end{equation}
\begin{equation}
J_C = 2.02 \times 10^{41} kg \: m^2 / sec  \; \; \; (\textrm{model III}).
\end{equation}	
Moving from one EoS to another induces a variation in $J$ of the order of 10\%, which should be sufficient to discriminate between different equation-of-states. 

\smallskip

\begin{figure*}[ht]
\centering
\includegraphics[width=0.32\textwidth]{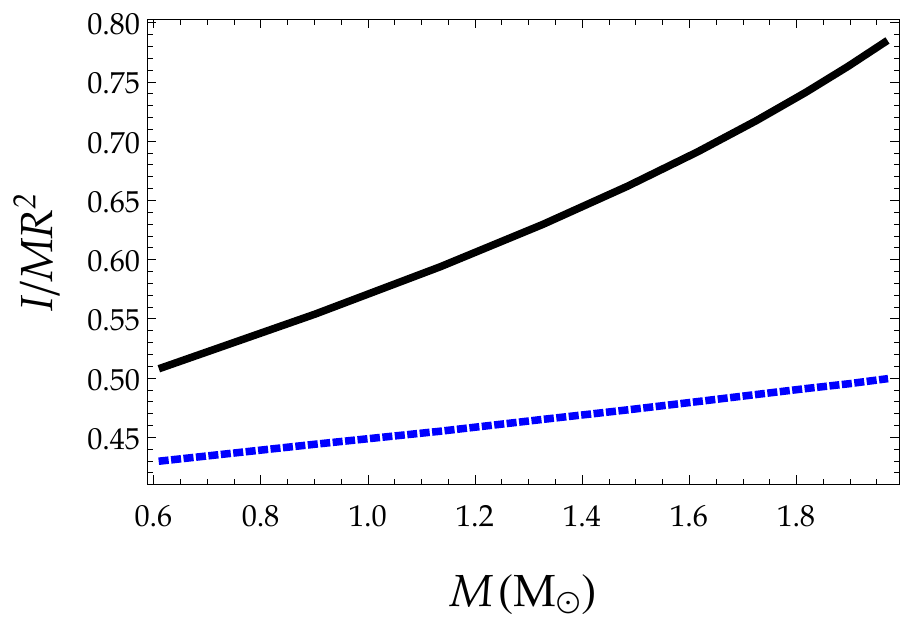}      \
\includegraphics[width=0.32\textwidth]{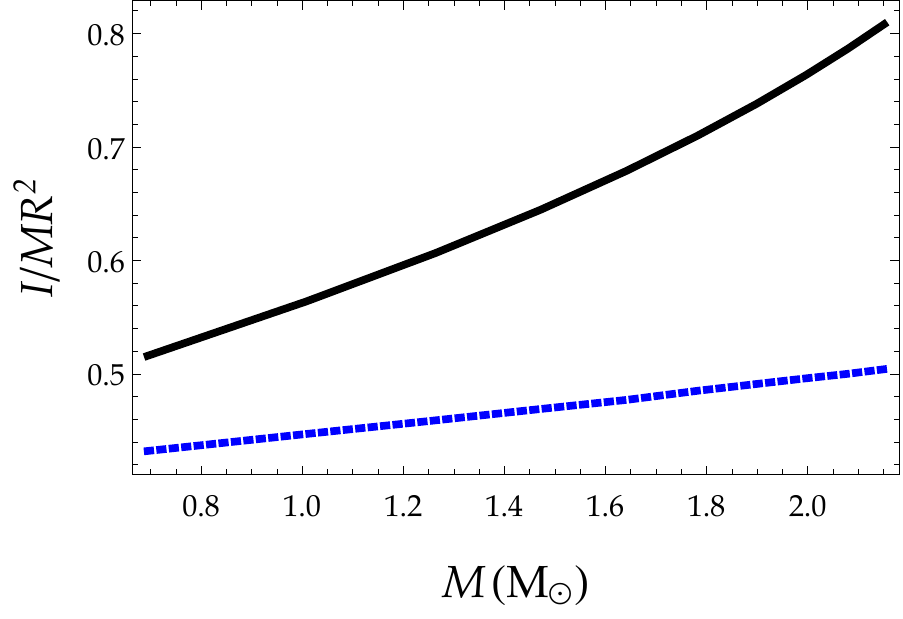}   \
\includegraphics[width=0.32\textwidth]{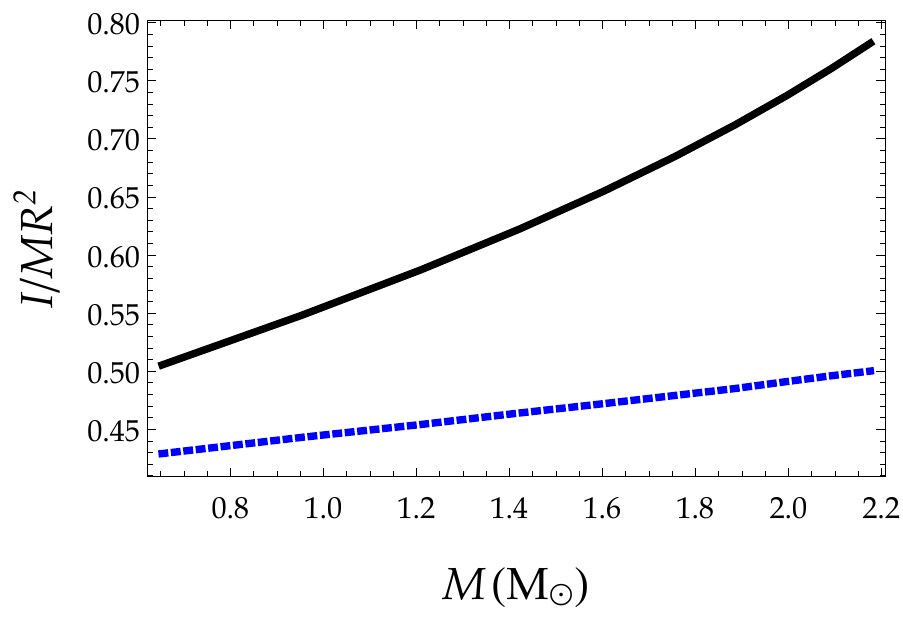}   \
\caption{
Dimensionless moment of inertia $I/(MR^2)$ as a function of the mass of the star (in solar masses) for different values of the parameters $\{A, B\}$.
{\bf{LEFT:}} panel corresponds to $A^2=0.4$ and $B=0.23 \times 10^{-3}/km^2$.
{\bf{MIDDLE:}} panel corresponds to $A^2=0.425$ and $B=0.215 \times 10^{-3} /km^2$.
{\bf{RIGHT:}} panel corresponds to $A^2=0.45$ and $B=0.2 \times 10^{-3} /km^2$.
}
\label{fig:1}
\end{figure*}

In Fig.~\ref{fig:1} we show the dimensionless moment of inertia $a \equiv I/(MR^2)$ against the mass of the compact object for models A (left panel), B (middle panel) and C (right panel). In each panel there are two curves, both for rotating and non-rotating stars for comparison reasons. The solid black line corresponds to the moment of inertia without rotation, whereas the dashed blue line corresponds to rotating stars. According to our results, we can make the the following observations: a) The moments of inertia increase with the mass of the star, b) in the case of non-rotating objects the moment of inertia grows faster, and c) the curve corresponding to rotation lies below the one corresponding to non-rotating stars. Therefore the deviation is smaller for light stars and larger for heavy stars. Moreover, for a given mass a rotating star has a lower moment of inertia than its non-rotating counterpart.

\smallskip

\begin{figure*}[ht]
\centering
\includegraphics[width=0.32\textwidth]{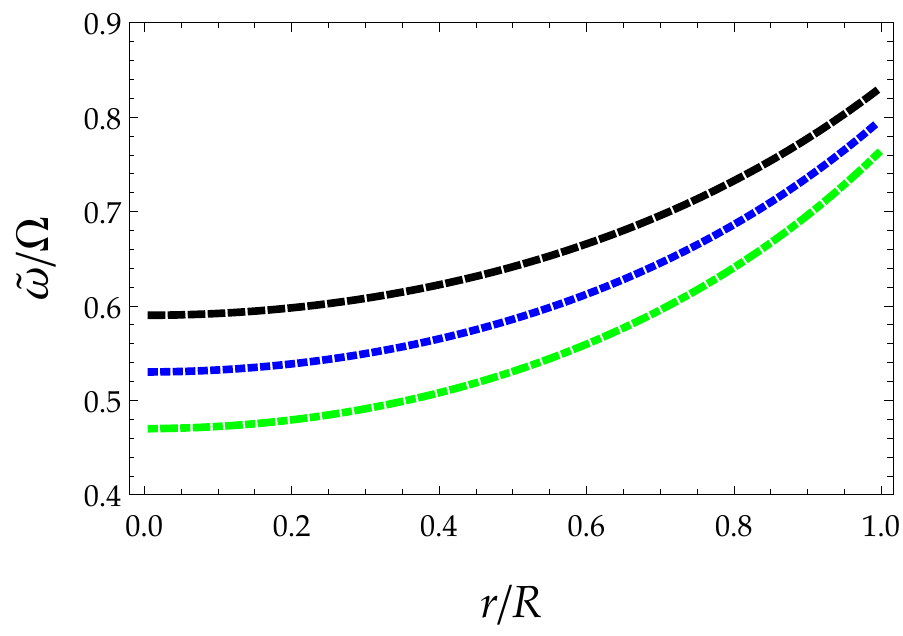}      \
\includegraphics[width=0.32\textwidth]{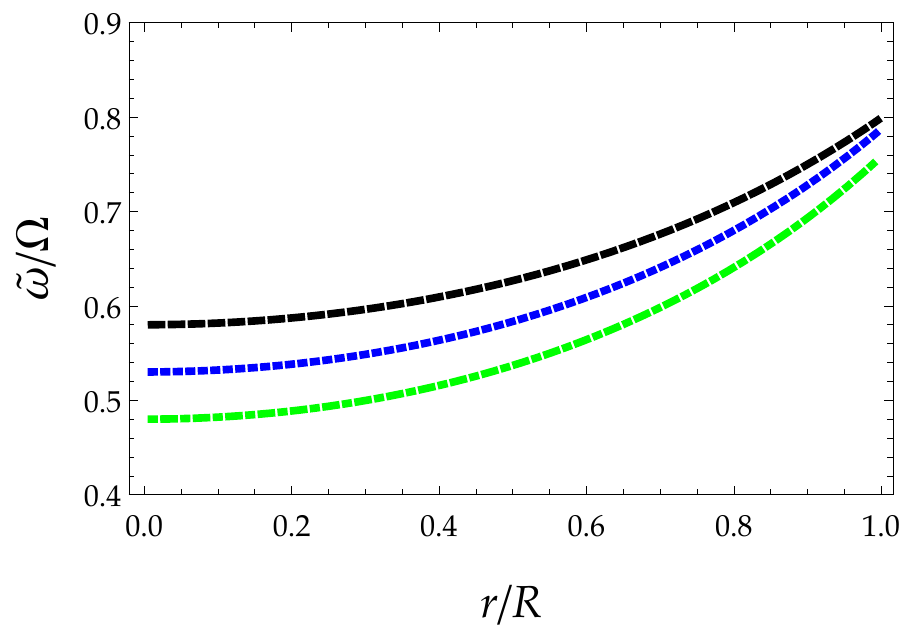}   \
\includegraphics[width=0.32\textwidth]{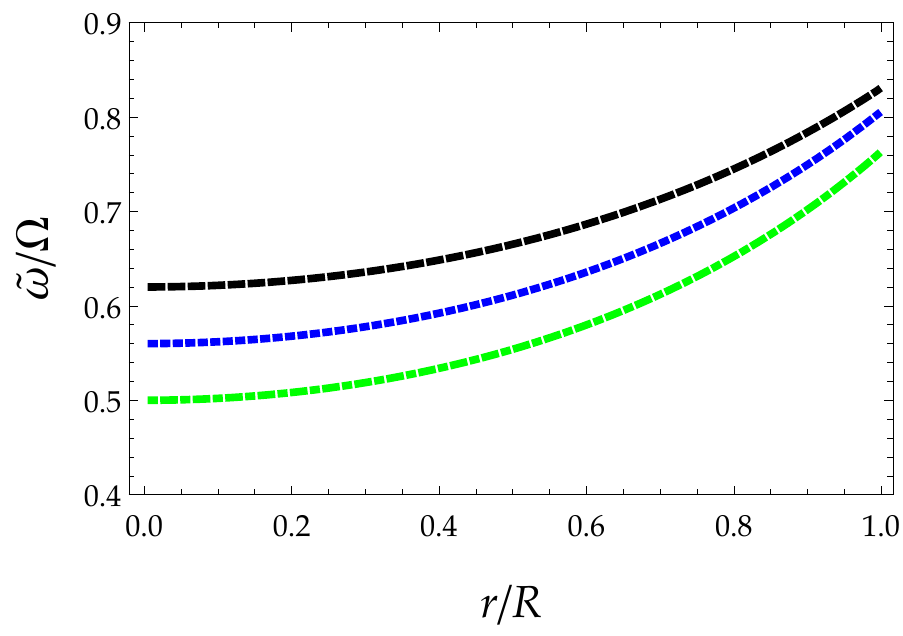}   \
\caption{
$\tilde{\omega}/\Omega$ vs  $r/R$ for different values of the parameters $\{A, B\}$ and for three masses aforementioned.
Each panel corresponds to a different set $\{A, B\}$, as in Fig.~1, while the three curves in each panel correspond to 
three different star masses. 
{\bf{LEFT:}} Shown are from top to bottom: $M=1.48~M_{\odot}$,  $M=1.73~M_{\odot}$,  $M=1.96~M_{\odot}$.
{\bf{MIDDLE:}} Shown are from top to bottom: $M=1.47~M_{\odot}$,  $M=1.78~M_{\odot}$,  $M=2.00~M_{\odot}$.
{\bf{RIGHT:}} Shown are from top to bottom: $M=1.42~M_{\odot}$,  $M=1.76~M_{\odot}$,  $M=2.00~M_{\odot}$.
}
\label{fig:2}
\end{figure*}

What is more, in Fig.~\ref{fig:2} we show the quantity $\tilde{\omega}/\Omega$ as a function of normalized radial coordinate, $r/R$,
for the three sets A, B and C. Each panel, corresponding to a different set, shows three solutions corresponding to three
different star masses as follows (from top to down): i) a light star with a mass around $~1.4~M_{\odot}$, ii) a star with an intermediate mass around $~1.75~M_{\odot}$, and iii) a heavy star with a mass around $~2~M_{\odot}$. The dashed black line represents a light star, dotted blued line an intermediate mass star, while dotted-dashed green line corresponds to a heavy star. In all cases $\tilde{\omega}/\Omega$ is an increasing function of $r/R$. Moreover, in all three panels as the mass of the object increases the curves are shifted downwards.

\smallskip

\begin{figure*}[ht]
\centering
\includegraphics[width=0.6\textwidth]{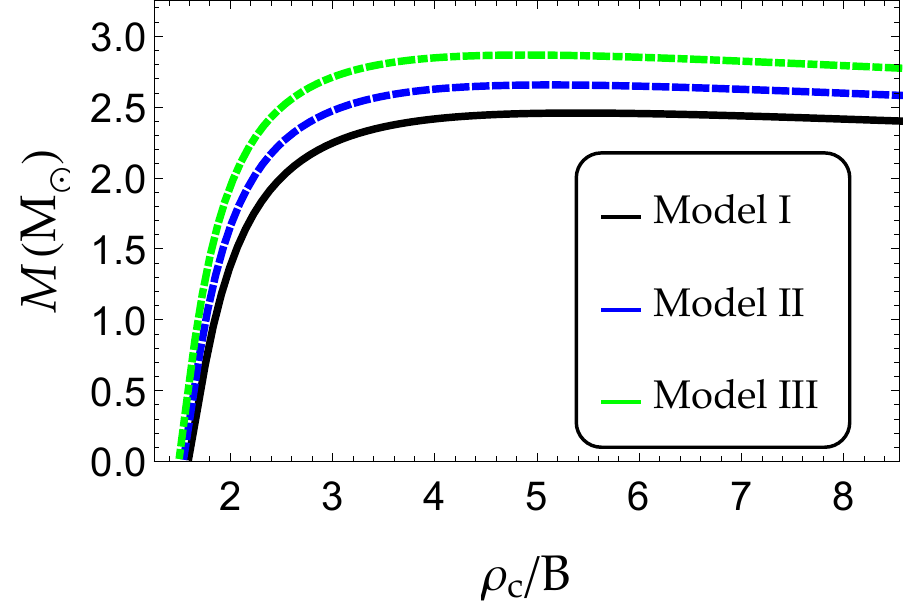}  
\caption{
Mass of the star versus (normalized) central energy density for the three models considered here.
}
\label{fig:3}
\end{figure*}

Before we conclude our work, a comment regarding stability is in order here. We now proceed to study the stellar mass $M$ against the central energy density $\rho_c$ for the three different models considered in the present work. We consider the so called static stability criterion \cite{harrison,ZN}
\begin{eqnarray}
&&\frac{d M}{d \rho_c} < 0 ~~~ \rightarrow \text{unstable configuration} \\
&&\frac{d M}{d \rho_c} > 0 ~~~ \rightarrow \text{stable configuration},
\end{eqnarray}
to be satisfied by all stellar configurations. According to Fig.~\ref{fig:3} the mass of the star reaches a maximum 
value at some $\rho_c^*$, and therefore the extremum point of the curve separates the stable from the unstable configuration. Figures~\ref{fig:1} and ~\ref{fig:2} refer to stable stars only.

%%%%%%%%%%%%%%%%%%%%%%%%%%%%%%BEGIN-FIGURES%%%%%%%%%%%%%%%%%%%%%%%%%%%%%%%%

%%%%%%%%%%%%%%%%%%%%%%%%%%%%%%%%%%%%%%%%%%%%%%%%%%%%%

%%%%%%%%%%%%%%%%%%%%%%%%%%%%%%%%%%%%%%%

%%%%%%%%%%%%%%%%%%%%%%%%%END-FIGURES%%%%%%%%%%%%%%%%%%%%%%%%%%%%%%%%%

%%%%%%%%%%%%%%%%%%%%%%
\section{Conclusions}
%%%%%%%%%%%%%%%%%%%%%%%

In summary, in the present work we have considered iso\-tropic and slowly-rotating dark energy stars, adopting the extended Chaplygin equation-of-state ("Chaplygin plus baro\-tropic"), characterized by two parameters $A,B$. The set of structure equations consists of the usual TOV equations for non-rotating stars plus one additional differential equation for the non-diagonal component of the metric tensor due to rotation. Due to slow rotation, all unknown quantities depend on the radial coordinate only. The complete system of coupled differential equations has been integrated numerically taking into account all the appropriate initial and boundary conditions. Finally, we have computed the moment of inertia as a function of the mass of the stars, both for rotating and non-rotating objects for comparison reasons. The solution for the non-diagonal metric component as a function of the radial coordinate is shown as well for three different star masses i) a light star ($M \sim 1.4 \ M_{\odot}$), ii) a heavy star ($M \sim 2 \ M_{\odot}$) and iii) an average star ($M \sim 1.75 \ M_{\odot}$). Our main findings may be summarized as follows: a) the moments of inertia increase with the mass of the star, b) in the case of non-rotating objects the moment of inertia grows faster, and c) the curve corresponding to rotation lies below the one corresponding to non-rotating stars. Therefore the deviation is smaller for light stars and larger for heavy stars. Moreover, for a given mass a rotating star has a lower moment of inertia than its non-rotating counterpart.

\smallskip

The NICER satellite is a NASA mission projected to observe thermal X-rays emitted by several millisecond pulsars. This
type of data could help us distinguish between the different EoS in such models. As pointed out in \cite{SudipBhattacharyya} a few years ago, the recent fast growth of millisecond pulsars with precisely measured mass provides us with an excellent opportunity to probe the physics of compact stars. Since the stellar parameter values can be computed accurately in two complementary scenarios, on the one hand, for known mass and spin rate, and on the other hand, for a chosen equation-of-state. The authors of \cite{SudipBhattacharyya} provided the first detailed catalog of numerically computed parameter values for 16 observed pulsars. Their study assumes eight different equations of state corresponding to nucleonic, hyperonic, hybrid and strange matter. The increase of observational data expected in the coming years will allow us to study the effect of rotation on the moments of inertia to validate or exclude this type of EoS models.

%%%%%%%%%%%%%%%%%%%%%%%%%%%%%%%%%%%%%%%%%%%%

\section*{Acknowlegements}

We thank the anonymous reviewer for useful comments and suggestions.
The authors G.~P. and I.~L. thank the Fun\-da\c c\~ao para a Ci\^encia e Tecnologia (FCT), Portugal, 
for the financial support to the Center for Astrophysics and Gravitation-CENTRA, Instituto Superior T\'ecnico, 
Universidade de Lisboa, through the Grants No. UID/FIS/00099/2013 and No. PTDC/FIS-AST/28920/2017. 
The author A.~R. acknowledges Universidad de Tarapac\'a for financial support.

%%%%%%%%%%%%%%%%%%%%%%%%%%%%%%%%%%%%%%%%%%%%


\begin{thebibliography}{99}
%
\bibitem{SN1} A.~G.~Riess et al. Astron. J. 116, 1009 (1998).

\bibitem{SN2} S.~Perlmutter et al., Astrophys. J. 517, 565 (1999).

\bibitem{turner} W.~L.~Freedman and M.~S.~Turner,
%``Measuring and understanding the universe,''
  Rev.\ Mod.\ Phys.\  {\bf 75} (2003) 1433
[astro-ph/0308418].

\bibitem{GR} A.~Einstein, 
  % "The Foundation of the General Theory of Relativity," 
Annalen Phys. 49 (1916) 769–822.

\bibitem{einstein} A.~Einstein,
  %``Cosmological Considerations in the General Theory of Relativity,''
  Sitzungsber.\ Preuss.\ Akad.\ Wiss.\ Berlin (Math.\ Phys.\ ) {\bf 1917} (1917) 142.
  
\bibitem{carroll} S.~M.~Carroll,
  %``The Cosmological constant,''
  Living Rev.\ Rel.\  {\bf 4} (2001) 1
%  doi:10.12942/lrr-2001-1
  [astro-ph/0004075].

\bibitem{weinberg} S.~Weinberg,
  %``The Cosmological Constant Problem,''
  Rev.\ Mod.\ Phys.\  {\bf 61} (1989) 1.
  
\bibitem{zeldovich} Y.~B.~Zeldovich,
  %``Cosmological Constant and Elementary Particles,''
  JETP Lett.\  {\bf 6} (1967) 316
  [Pisma Zh.\ Eksp.\ Teor.\ Fiz.\  {\bf 6} (1967) 883].


\bibitem{tension} B.~Ryden,
  %``A constant conflict,''
  Nature Phys.\  {\bf 13} (2017) no.3,  314.

\bibitem{tension1} L.~Verde, P.~Protopapas and R.~Jimenez,
  %``Planck and the local Universe: Quantifying the tension,''
  Phys.\ Dark Univ.\  {\bf 2} (2013) 166
  [arXiv:1306.6766 [astro-ph.CO]].

\bibitem{tension2} K.~Bolejko,
  %``Emerging spatial curvature can resolve the tension between high-redshift CMB and low-redshift distance ladder measurements of the Hubble constant,''
  Phys.\ Rev.\ D {\bf 97} (2018) no.10,  103529
 % [arXiv:1712.02967 [astro-ph.CO]].

\bibitem{tension3} E.~Mörtsell and S.~Dhawan,
  %``Does the Hubble constant tension call for new physics?,''
  arXiv:1801.07260 [astro-ph.CO].
  
\bibitem{planck1} P.~A.~R.~Ade {\it et al.} [Planck Collaboration],
  %``Planck 2015 results. XIII. Cosmological parameters,''
  Astron.\ Astrophys.\  {\bf 594} (2016) A13
  [arXiv:1502.01589 [astro-ph.CO]].
  
\bibitem{planck2} N.~Aghanim {\it et al.} [Planck Collaboration],
  %``Planck 2018 results. VI. Cosmological parameters,''
  arXiv:1807.06209 [astro-ph.CO].
  
\bibitem{hubble} A.~G.~Riess {\it et al.},
  %``A 2.4% Determination of the Local Value of the Hubble Constant,''
  Astrophys.\ J.\  {\bf 826} (2016) no.1,  56
  [arXiv:1604.01424 [astro-ph.CO]].
  
\bibitem{recent} A.~G.~Riess {\it et al.},
  %``Milky Way Cepheid Standards for Measuring Cosmic Distances and Application to Gaia DR2: Implications for the Hubble Constant,''
  Astrophys.\ J.\  {\bf 861} (2018) no.2,  126
  [arXiv:1804.10655 [astro-ph.CO]].  

\bibitem{newphysics} E.~M{\"o}rtsell and S.~Dhawan,
  %``Does the Hubble constant tension call for new physics?,''
  JCAP {\bf 1809} (2018) no.09,  025
%  doi:10.1088/1475-7516/2018/09/025
  [arXiv:1801.07260 [astro-ph.CO]].
  
  
\bibitem{mod1} T.~P.~Sotiriou and V.~Faraoni,
  %``f(R) Theories Of Gravity,''
  Rev.\ Mod.\ Phys.\  {\bf 82} (2010) 451
  [arXiv:0805.1726 [gr-qc]].

\bibitem{mod2} A.~De Felice and S.~Tsujikawa,
  %``f(R) theories,''
  Living Rev.\ Rel.\  {\bf 13} (2010) 3
  [arXiv:1002.4928 [gr-qc]].

\bibitem{HS} W.~Hu and I.~Sawicki,
  %``Models of f(R) Cosmic Acceleration that Evade Solar-System Tests,''
  Phys.\ Rev.\ D {\bf 76} (2007) 064004
  [arXiv:0705.1158 [astro-ph]].

\bibitem{starobinsky} A.~A.~Starobinsky,
  %``Disappearing cosmological constant in f(R) gravity,''
  JETP Lett.\  {\bf 86} (2007) 157
 %[arXiv:0706.2041 [astro-ph]].
 
\bibitem{langlois} D.~Langlois,
  %``Brane cosmology: An Introduction,''
  Prog.\ Theor.\ Phys.\ Suppl.\  {\bf 148} (2003) 181
  [hep-th/0209261].

\bibitem{maartens} R.~Maartens,
  %``Brane world gravity,''
  Living Rev.\ Rel.\  {\bf 7} (2004) 7
  [gr-qc/0312059].

\bibitem{dgp} G.~R.~Dvali, G.~Gabadadze and M.~Porrati,
  %``4-D gravity on a brane in 5-D Minkowski space,''
  Phys.\ Lett.\ B {\bf 485} (2000) 208
[hep-th/0005016].

\bibitem{BD1} C.~Brans and R.~H.~Dicke,
  %``Mach's principle and a relativistic theory of gravitation,''
  Phys.\ Rev.\  {\bf 124} (1961) 925.
  
\bibitem{BD2} C.~H.~Brans,
  %``Mach's Principle and a Relativistic Theory of Gravitation. II,''
  Phys.\ Rev.\  {\bf 125}, 2194 (1962).
  
\bibitem{leandros} J.~C.~B.~Sanchez and L.~Perivolaropoulos,
  %``Evolution of Dark Energy Perturbations in Scalar-Tensor Cosmologies,''
  Phys.\ Rev.\ D {\bf 81} (2010) 103505
[arXiv:1002.2042 [astro-ph.CO]].

\bibitem{PR} G.~Panotopoulos and \'A.~Rinc\'on,
  %``Stability of cosmic structures in scalar–tensor theories of gravity,''
  Eur.\ Phys.\ J.\ C {\bf 78} (2018) no.1,  40
  [arXiv:1710.02485 [astro-ph.CO]].
  
\bibitem{DE1} B.~Ratra and P.~J.~E.~Peebles,
  %``Cosmological Consequences of a Rolling Homogeneous Scalar Field,''
  Phys.\ Rev.\ D {\bf 37} (1988) 3406.

\bibitem{DE2} I.~Y.~Aref'eva, A.~S.~Koshelev and S.~Y.~Vernov,
  %``Exact solution in a string cosmological model,''
  Theor.\ Math.\ Phys.\  {\bf 148} (2006) 895
   [Teor.\ Mat.\ Fiz.\  {\bf 148} (2006) 23]
  [astro-ph/0412619].

\bibitem{DE3} R.~Lazkoz and G.~Leon,
  %``Quintom cosmologies admitting either tracking or phantom attractors,''
  Phys.\ Lett.\ B {\bf 638} (2006) 303
  [astro-ph/0602590].

\bibitem{DE4} J.~S.~Bagla, H.~K.~Jassal and T.~Padmanabhan,
  %``Cosmology with tachyon field as dark energy,''
  Phys.\ Rev.\ D {\bf 67} (2003) 063504
  [astro-ph/0212198].

\bibitem{DE5} C.~Armendariz-Picon, V.~F.~Mukhanov and P.~J.~Steinhardt,
  %``Essentials of k essence,''
  Phys.\ Rev.\ D {\bf 63} (2001) 103510
  [astro-ph/0006373].  
  
\bibitem{copeland} E.~J.~Copeland, M.~Sami and S.~Tsujikawa,
%``Dynamics of dark energy,''
  Int.\ J.\ Mod.\ Phys.\ D {\bf 15} (2006) 1753
[hep-th/0603057].  

\bibitem{Chaplygin1} A.~Y.~Kamenshchik, U.~Moschella and V.~Pasquier,
%``An Alternative to quintessence,''
Phys. Lett. B \textbf{511} (2001), 265-268
%doi:10.1016/S0370-2693(01)00571-8
[arXiv:gr-qc/0103004 [gr-qc]].

\bibitem{Chaplygin2} M.~C.~Bento, O.~Bertolami and A.~A.~Sen,
%``Generalized Chaplygin gas, accelerated expansion and dark energy matter unification,''
Phys. Rev. D \textbf{66} (2002), 043507
%doi:10.1103/PhysRevD.66.043507
[arXiv:gr-qc/0202064 [gr-qc]].


\bibitem{Tulin} S.~Tulin and H.~B.~Yu,
%``Dark Matter Self-interactions and Small Scale Structure,''
Phys. Rept. \textbf{730} (2018), 1-57
%doi:10.1016/j.physrep.2017.11.004
[arXiv:1705.02358 [hep-ph]].

\bibitem{DMS1} X.~Y.~Li, T.~Harko and K.~S.~Cheng,
%``Condensate dark matter stars,''
JCAP \textbf{06} (2012), 001
%doi:10.1088/1475-7516/2012/06/001
[arXiv:1205.2932 [astro-ph.CO]].

\bibitem{DMS2} A.~Maselli, P.~Pnigouras, N.~G.~Nielsen, C.~Kouvaris and K.~D.~Kokkotas,
%``Dark stars: gravitational and electromagnetic observables,''
Phys. Rev. D \textbf{96} (2017) no.2, 023005
%doi:10.1103/PhysRevD.96.023005
[arXiv:1704.07286 [astro-ph.HE]].

\bibitem{DMS3} G.~Panotopoulos and I.~Lopes,
%``Dark stars in Starobinsky’s model,''
Phys. Rev. D \textbf{97} (2018) no.2, 024025
%doi:10.1103/PhysRevD.97.024025
[arXiv:1801.03387 [gr-qc]].

\bibitem{DMS4} A.~Maselli, C.~Kouvaris and K.~D.~Kokkotas,
  %``Photon spectrum of asymmetric dark stars,''
  Int.\ J.\ Mod.\ Phys.\ D {\bf 30} (2021) no.01,  2150003
  doi:10.1142/S0218271821500036
  [arXiv:1905.05769 [astro-ph.CO]].

\bibitem{exotic} K.~N.~Singh, A.~Ali, F.~Rahaman and S.~Nasri,
%``Compact stars with exotic matter,''
Phys. Dark Univ. \textbf{29} (2020), 100575
%doi:10.1016/j.dark.2020.100575
[arXiv:2005.00540 [gr-qc]].

\bibitem{paperbase1} F.~Tello-Ortiz, M.~Malaver, \'A.~Rinc{\'o}n and Y.~Gomez-Leyton,
%``Relativistic anisotropic fluid spheres satisfying a non-linear equation of state,''
Eur. Phys. J. C \textbf{80} (2020) no.5, 371
%doi:10.1140/epjc/s10052-020-7956-0
[arXiv:2005.11038 [gr-qc]].

\bibitem{paperbase2} G.~Panotopoulos, A.~Rincon and I.~Lopes,
  %``Radial oscillations and tidal Love numbers of dark energy stars,''
  Eur.\ Phys.\ J.\ Plus {\bf 135} (2020) no.10,  856
%  doi:10.1140/epjp/s13360-020-00867-x
  [arXiv:2010.09373 [gr-qc]].
  

\bibitem{herre} L. Herrera and N. O. Santos, 
\emph{Phys. Rep.} \textbf{286}, 53 (1997).

\bibitem{harko} M. K. Mak and T. Harko,  \emph{Proc. Roy. Soc. Lond. A} \textbf{459}, 393 (2003).

\bibitem{a2} M. Cosenza, L. Herrera, M. Esculpi and L. Witten, 
\emph{Phys. Rev. D} \textbf{a3}, 2527 (1982).

\bibitem{a4} L. Herrera and J. Ponce de Le\'on, \emph{J. Math. Phys.} \textbf{26}, 2302 (1985).

\bibitem{a5} J. Ponce de Le\'on, \emph{Gen. Relativ. Gravit.} \textbf{19}, 797 (1987).

\bibitem{a6} J. Ponce de Le\'on, \emph{J. Math. Phys.} \textbf{28}, 1114 (1987).

\bibitem{a7} R. Chan, S. Kichenassamy, G. Le Denmat and  N. O. Santos, \emph{Mon. Not. R. Astron. Soc.} \textbf{239}, 91 (1989).

\bibitem{a8} H. Bondi, \emph{Mon. Not. R. Astron. Soc.} \textbf{259}, 365 (1992).

\bibitem{a9} R. Chan, L. Herrera and N. O. Santos, \emph{Class. Quantum Grav.} \textbf{9}, 133 (1992).

\bibitem{a10} R. Chan, L. Herrera and N. O. Santos, \emph{Mon. Not. R. Astron. Soc.} \textbf{265}, 533 (1993).

\bibitem{a11} M. K. Gokhroo and A. L. Mehra,  \emph{Gen. Rel. Grav.} \textbf{26}, 75 (1994).

\bibitem{a12} A. Di Prisco,  E. Fuenmayor, L. Herrera, V. Varela, \emph{Phys. Lett. A} \textbf{195}, 23 (1994).

\bibitem{a13} A. Di Prisco, L. Herrera and V. Varela, \emph{Gen. Relativ. Gravit.} \textbf{29}, 1239 (1997).

\bibitem{a15} K. Dev and M. Gleiser, Gen. Relativ. Gravit. 34, 1793 (2002).

\bibitem{a16} M. K. Mak and T. Harko, \emph{Chin. J. Astron. Astrophys.} \textbf{2}, 248 (2002).

\bibitem{a17} M. K. Mak, P. N. Dobson and T. Harko, \emph{Int. J. Mod. Phys. D} \textbf{11}, 207 (2002).

\bibitem{a19} H. Abreu, H. Hern\'andez and L. A. N\'u\~nez, \emph{Class. Quantum. Grav.} \textbf{24}, 4631 (2007).

\bibitem{a20} S. Viaggiu, \emph{Int. J. Mod. Phys.D} \textbf{18}, 275 (2009).

\bibitem{a21} R. P. Negreiros, F. Weber, M. Malheiro and V. Usov, \emph{Phys. Rev. D} \textbf{80} 083006, (2009).

\bibitem{a22} B.V. Ivanov, \emph{Int. J. Theor. Phys.} \textbf{49}, 1236 (2010).

\bibitem{Panotopoulos:2021obe} G.~Panotopoulos, A.~Rincon and I.~Lopes,
  %``Interior solutions of relativistic stars with anisotropic matter in scale-dependent gravity,''
  Eur.\ Phys.\ J.\ C {\bf 81}, no. 1, 63 (2021)
  %doi:10.1140/epjc/s10052-021-08881-8
  [arXiv:2101.06649 [gr-qc]].

\bibitem{Panotopoulos:2020kgl} G.~Panotopoulos, A.~Rincon and I.~Lopes,
  %``Radial oscillations and tidal Love numbers of dark energy stars,''
  Eur.\ Phys.\ J.\ Plus {\bf 135}, no. 10, 856 (2020)
%  doi:10.1140/epjp/s13360-020-00867-x
  [arXiv:2010.09373 [gr-qc]].

\bibitem{Bhar:2020ukr} P.~Bhar, F.~Tello-Ortiz, A.~Rincon and Y.~Gomez-Leyton,
  %``Study on anisotropic stars in the framework of Rastall gravity,''
  Astrophys.\ Space Sci.\  {\bf 365}, no. 8, 145 (2020).
 % doi:10.1007/s10509-020-03859-6

\bibitem{Abellan:2020dze} G.~Abellán, A.~Rincon, E.~Fuenmayor and E.~Contreras,
  %``Anisotropic interior solution by gravitational decoupling based on a non-standard anisotropy,''
  Eur.\ Phys.\ J.\ Plus {\bf 135}, no. 7, 606 (2020).
%  doi:10.1140/epjp/s13360-020-00589-0

\bibitem{Tello-Ortiz:2020nuc} F.~Tello-Ortiz, A.~Rincon, P.~Bhar and Y.~Gomez-Leyton,
  %``Durgapal IV model in light of the minimal geometric deformation approach,''
  Chin.\ Phys.\ C {\bf 44}, 105102 (2020)
 % doi:10.1088/1674-1137/aba5f7
  [arXiv:2006.04512 [gr-qc]].

\bibitem{Tello-Ortiz:2020svg} F.~Tello-Ortiz, M.~Malaver, A.~Rincon and Y.~Gomez-Leyton,
  %``Relativistic anisotropic fluid spheres satisfying a non-linear equation of state,''
  Eur.\ Phys.\ J.\ C {\bf 80}, no. 5, 371 (2020)
  %doi:10.1140/epjc/s10052-020-7956-0
  [arXiv:2005.11038 [gr-qc]].

\bibitem{Panotopoulos:2020zqa} G.~Panotopoulos, A.~Rincon and I.~Lopes,
  %``Interior solutions of relativistic stars in the scale-dependent scenario,''
  Eur.\ Phys.\ J.\ C {\bf 80}, no. 4, 318 (2020)
 % doi:10.1140/epjc/s10052-020-7900-3
  [arXiv:2004.02627 [gr-qc]].

\bibitem{Panotopoulos:2019zxv} G.~Panotopoulos and A.~Rincon,
  %``Relativistic strange quark stars in Lovelock gravity,''
  Eur.\ Phys.\ J.\ Plus {\bf 134}, no. 9, 472 (2019)
 % doi:10.1140/epjp/i2019-12853-1
  [arXiv:1907.03545 [gr-qc]].

\bibitem{Gabbanelli:2018bhs} L.~Gabbanelli, Á.~Rincón and C.~Rubio,
  %``Gravitational decoupled anisotropies in compact stars,''
  Eur.\ Phys.\ J.\ C {\bf 78}, no. 5, 370 (2018)
  %doi:10.1140/epjc/s10052-018-5865-2
  [arXiv:1802.08000 [gr-qc]].

\bibitem{Panotopoulos:2021cxu} G.~Panotopoulos, T.~Tangphati and A.~Banerjee,
  %``Electrically charged compact stars with an interacting quark equation of state,''
  arXiv:2105.10638 [gr-qc].

\bibitem{Panotopoulos:2021sbf} G.~Panotopoulos, T.~Tangphati, A.~Banerjee and M.~K.~Jasim,
  %``Anisotropic quark stars in $R^2$ gravity,''
  Phys.\ Lett.\ B {\bf 817}, 136330 (2021)
 % doi:10.1016/j.physletb.2021.136330
  [arXiv:2104.00590 [gr-qc]].

\bibitem{Moraes:2021lhh} P.~H.~R.~S.~Moraes, G.~Panotopoulos and I.~Lopes,
  %``Anisotropic Dark Matter Stars,''
  Phys.\ Rev.\ D {\bf 103}, no. 8, 084023 (2021)
%  doi:10.1103/PhysRevD.103.084023
  [arXiv:2101.02207 [gr-qc]].

\bibitem{Tello-Ortiz:2020euy} F.~Tello-Ortiz, S.~K.~Maurya and Y.~Gomez-Leyton,
  %``Class I approach as MGD generator,''
  Eur.\ Phys.\ J.\ C {\bf 80}, no. 4, 324 (2020).
 % doi:10.1140/epjc/s10052-020-7882-1

\bibitem{Tello-Ortiz:2019gcl} F.~Tello-Ortiz, S.~K.~Maurya, A.~Errehymy, K.~N.~Singh and M.~Daoud,
  %``Anisotropic relativistic fluid spheres: an embedding class I approach,''
  Eur.\ Phys.\ J.\ C {\bf 79}, no. 11, 885 (2019).
  
\bibitem{ExtCh1} B.~Pourhassan and E.~O.~Kahya,
  %``Extended Chaplygin gas model,''
  Results Phys.\  {\bf 4} (2014) 101.

\bibitem{ExtCh2} V.~M.~C.~Ferreira and P.~P.~Avelino,
  %``Extended family of generalized Chaplygin gas models,''
  Phys.\ Rev.\ D {\bf 98} (2018) no.4,  043515
%  doi:10.1103/PhysRevD.98.043515
  [arXiv:1807.04656 [gr-qc]].
  
\bibitem{Extra1} E.~O.~Kahya and B.~Pourhassan,
  %``Observational constraints on the extended Chaplygin gas inflation,''
  Astrophys.\ Space Sci.\  {\bf 353} (2014) no.2,  677.

\bibitem{Extra2} B.~Pourhassan,
  %``Extended Chaplygin gas in Horava–Lifshitz gravity,''
  Phys.\ Dark Univ.\  {\bf 13} (2016) 132
%  doi:10.1016/j.dark.2016.06.002
  [arXiv:1412.2605 [gr-qc]].

\bibitem{Extra3} B.~Pourhassan,
  %``Unified universe history through phantom extended Chaplygin gas,''
  Can.\ J.\ Phys.\  {\bf 94} (2016) no.7,  659
%  doi:10.1139/cjp-2016-0154
  [arXiv:1504.04173 [gr-qc]].
    
\bibitem{Tolman} R.~C.~Tolman,
  %``Static solutions of Einstein's field equations for spheres of fluid,''
  Phys.\ Rev.\  {\bf 55}, 364 (1939).
  %doi:10.1103/PhysRev.55.364
  %%CITATION = doi:10.1103/PhysRev.55.364;%%
  %748 citations counted in INSPIRE as of 15 Jun 2017
   
\bibitem{OV} J.~R.~Oppenheimer and G.~M.~Volkoff,
  %``On Massive neutron cores,''
Phys.\ Rev.\  {\bf 55} (1939) 374.   

\bibitem{Gourgoulhon:2010ju} E.~Gourgoulhon,
%``An Introduction to the theory of rotating relativistic stars,''
[arXiv:1003.5015 [gr-qc]].

\bibitem{Paschalidis:2016vmz} V.~Paschalidis and N.~Stergioulas,
%``Rotating Stars in Relativity,''
Living Rev. Rel. \textbf{20} (2017) no.1, 7
%doi:10.1007/s41114-017-0008-x
[arXiv:1612.03050 [astro-ph.HE]].  

\bibitem{Hartle:1967he} J.~B.~Hartle,
%``Slowly rotating relativistic stars. 1. Equations of structure,''
Astrophys. J. \textbf{150} (1967), 1005-1029

\bibitem{Staykov:2014mwa} K.~V.~Staykov, D.~D.~Doneva, S.~S.~Yazadjiev and K.~D.~Kokkotas,
%``Slowly rotating neutron and strange stars in $R^2$ gravity,''
JCAP \textbf{10} (2014), 006
%doi:10.1088/1475-7516/2014/10/006
[arXiv:1407.2180 [gr-qc]].

\bibitem{PLPulsars} G.~Panotopoulos and I.~Lopes,
  %``Millisecond pulsars modeled as strange quark stars admixed with condensed dark matter,''
  Int.\ J.\ Mod.\ Phys.\ D {\bf 27} (2018) no.09,  1850093
%  doi:10.1142/S0218271818500931
  [arXiv:1804.05023 [gr-qc]].

  
\bibitem{r19} L. Xu, J. Lu and Y Wang, \emph{Eur. Phys. J. C} \textbf{72}, 1883 (2012). 

\bibitem{r20} H. Saadat H and B. Pourhassan, \emph{Astrophys. Space Sci.} \textbf{343}, 783 (2013).

\bibitem{r21} H. Saadat and B. Pourhassan, \emph{Astrophys Space Sci.} \textbf{344}, 237 (2013). 

\bibitem{r22} B. Pourhassan, \emph{Int. J. Modern Phys. D} \textbf{22}, 1350061 (2013).

\bibitem{r26} A. Chanda, S. Dey and B. C. Paul, \emph{Eur. Phys. J. C} \textbf{79}, 502 (2019).

\bibitem{r27} P. Saha and U. Debnath, \emph{Eur. Phys. J. C} \textbf{79}, 919 (2019).

\bibitem{r28} F. S. N. Lobo, \emph{Phys.Rev. D} \textbf{73},  064028 (2006).

\bibitem{r25} P. Bhar and M. Govender, \emph{Int. J. Modern Phys. D} \textbf{26}, 1750053 (2017).

\bibitem{r29} M. Salti, O. Aydogdu, H. Yanar and K. Sogut, \emph{Ann. of Phys.} \textbf{390}, 131 (2018).

\bibitem{SudipBhattacharyya} F.~Tello-Ortiz, S.~K.~Maurya, A.~Errehymy, K.~N.~Singh and M.~Daoud,
S.~Bhattacharyya, I.~Bombaci, D.~Bandyopadhyay, A.~V.~Thampan, D.~Logoteta,
%``Millisecond radio pulsars with known masses: parameter values and equation of state models,''
New Astronomy, {\bf 54} , p. 61-71 (2017).

\bibitem{harrison} B.~K.~Harrison, \textit{Gravitational Theory and Gravitational Collapse}, University of Chicago Press, Chicago, 1965.

\bibitem{ZN} Y.~B.~Zeldovich, and I.~D.~Novikov, \textit{Relativistic Astrophysics, Vol.~I: Stars and Relativity}, University of Chicago Press, Chicago, 1971.
%
\end{thebibliography}
\end{document}